\documentclass{llncs}

\usepackage{geometry}
\usepackage{lipsum}

\usepackage{dblfloatfix}
\usepackage{lscape}
\usepackage{amsmath}
\usepackage{url}
\usepackage{tabularx}
\usepackage{multirow}
\usepackage{caption}
\usepackage{subcaption}
\usepackage{color, colortbl}
\usepackage{booktabs}
\usepackage{comment}
\usepackage{float}
\usepackage{fixltx2e}

\usepackage{multirow}

\usepackage{subfig}

\usepackage[table]{xcolor}

  \usepackage[pdftex]{graphicx}

\newcommand{\EG}[1]{EG$_{\text{#1}}$}

\newcommand{\ourDelta}[2][]{\ensuremath{\Delta\textsuperscript{\text{#1}}
\textsubscript{\text{#2}}}}

\newcommand{\high}[1][ ]{\ensuremath{\text{\emph{high}}#1}}
\newcommand{\low}[1][ ]{\ensuremath{\text{\emph{low}}#1}}

\newcommand{\highDelta}[2][]{\high{\ourDelta[#1]{#2}}}
\newcommand{\lowDelta}[2][]{\low{\ourDelta[#1]{#2}}}

\newcommand{\threshold}[1][]{\ensuremath{\tau\textsubscript{#1}}}
\newcommand{\symptom}[1][]{\ensuremath{\sigma\textsubscript{#1}}}

\begin{document}
%
\title{A Centralized Mechanism to Make Predictions \\ Based on Data From Multiple 
WSNs}

\author{Gabriel Martins Dias \and Simon Oechsner \and 
Boris Bellalta}

\authorrunning{Gabriel Martins Dias et al.} 

%
\tocauthor{Gabriel Martins Dias, Simon Oechsner, and 
Boris Bellalta}

\institute{Department of Information and Communication Technologies \\ Pompeu Fabra University, Barcelona, Spain}

\maketitle

\begin{abstract}


In this work, we present a method that exploits a scenario with inter-Wireless 
Sensor Networks (WSNs) information exchange by making predictions and adapting 
the workload of a WSN according to their outcomes. We show the feasibility of 
an approach that intelligently utilizes information produced by other WSNs that 
may or not belong to the same administrative domain. To illustrate how the 
predictions using data from external WSNs can be utilized, a specific use-case 
is considered, where the operation of a WSN measuring relative humidity is 
optimized using the data obtained from a WSN measuring temperature. 
Based on a dedicated performance score, the simulation results show that this 
new approach can find the optimal operating point associated to the trade-off 
between energy consumption and quality of measurements. Moreover, we outline the 
additional challenges that need to be overcome, and draw conclusions to guide 
the future work in this field.

\end{abstract}

\section{Introduction}

Nowadays, forests, cities and houses, among others, are monitored by multiple 
Wireless Sensor Networks (WSNs) that may belong to different organizations, both 
public and private, as well as to individual citizens. In addition, there is a 
high heterogeneity regarding the technologies, protocols and standards used in 
WSNs. In this situation, each WSN usually operates completely independent of 
other WSNs, even if they are covering the same physical area, and is thus not 
able to take any advantage of the presence of those other WSNs to enrich its 
collected data nor to optimize its operation.

However, WSN performance can be improved by combining data generated from 
different sensors, belonging to the same node, other nodes from the same network 
or from other WSNs. This data sharing allows each WSN to build a deeper 
knowledge about its surroundings, may reduce the probability of getting wrong 
values and taking wrong decisions, and encompasses wider areas and different 
perspectives of the same environment. 


In an era of high availability of data from the cloud, we are interested in 
using data from other WSNs to reduce the energy consumption and improve the 
quality of the measurements done by a target WSN. The external information will 
be used to make predictions and change the operation of the nodes and save 
energy when the environmental conditions do not indicate that big changes will 
happen in the near future. For example, relative humidity and temperature values 
usually have a high correlation, and the former may have a higher variation if 
the latter is changing.

This paper lists some of the existing alternatives for collaboration and 
prediction in WSNs and develops further the inter-WSNs information exchange 
concept introduced in \cite{Pal2012} and in \cite{6583430}. The main idea behind 
the inter-WSN information exchange is that the data gathered by other WSNs can 
be exchanged via their sinks and used to improve the operation of the target 
one, and vice versa. 
Our main contribution is a mechanism that uses the data from collaborating WSNs 
to make predictions.
In order to validate our idea, we show how the WSNs evolve using this kind of 
collaboration, define a way to scale the quality of the measurements and the 
WSNs' performance, and finally present some simulation results from a chosen 
scenario consisting of two WSNs, one for monitoring the relative humidity and 
another for the temperature. Based on the presented results, we show how 
energy-efficient and accurate it can be.

The paper is organized in the following sections: In 
Section~\ref{sec:related-work}, we describe related works about collaboration 
between WSNs, the use of data from external sensors and predictions in WSN 
environments; the details of our proposed mechanism are explained in 
Section~\ref{sec:simulations}; the use case considered for the tests is 
detailed in Section~\ref{sec:current-simulations-scenario}; the simulation 
results and the evaluation of the approach are explained in 
Section~\ref{sec:evaluation} and; at the end, our conclusions and ideas for 
future work are shown in Section~\ref{sec:conclusion}.

\section{Related Work}
\label{sec:related-work}

A system that combines the action of individual components may produce better 
results than the individual components acting separately. Supported by this 
premisse, several collaboration mechanisms in WSNs have been developed.
Most of the approaches explore the collaboration between sensor nodes of the 
same WSN. In contrast to them, we extend the concept of collaboration to an 
upper layer and build the information exchange between different WSNs, without 
losing any other possible collaboration from the other levels. 

An inter-domain routing protocol is described in~\cite{Dressler2010}, where it 
is shown that the gateways may share information about their nodes and take 
advantage of being physically close to each other. This information can be used 
to transmit packets through nodes of the other WSNs and can be done either 
to share the information or for routing purposes. Even though the idea of our 
work is to create a link between nodes from different WSNs, it is neither meant 
to share resources nor information between wireless sensor nodes, but the 
knowledge that the network is able to produce based on collected data.

In \cite{Parker2008}, the authors describe a scenario where a system 
is responsible for building a richer knowledge about the environment by making 
use of the information produced by other WSNs. In their example, wireless 
sensor nodes combine sensory information with their localization and help 
other systems to localize and track objects from a distance. The goal of the 
described approach is to enable a robot to use the data retrieved by a WSN that 
detects the presence of objects inside the monitored area. After receiving the 
information from the WSN, the robot interprets the position of the object and 
moves itself to its location in order to get more details about the real 
situation. Their approach is different from ours mainly because it uses a 
non-generic solution that is highly coupled to the presented scenario without a 
WSN as the beneficiary of the collected information, besides not making any 
prediction with the information received from the others.

Besides the works that encourage the collaboration among WSNs, some authors 
applied predictions in order to reduce the energy consumption in the WSNs and 
extend their lifetime.
In~\cite{Yann-Ael2005}, the authors developed an algorithm for WSN applications 
that require a continuous delivery of sensor measurements, such as temperature 
and traffic monitoring. In order to build sets of nodes that provide trustful 
measurements, it considers that a sensor measurement is predictable if the 
predicted value (on average) differs on less than a (user) defined threshold 
when using other nodes' measurements.
After defining which sensors can be predicted by which other, the base station 
must find a set of subsets of active nodes such that a different prediction 
subset is used at each time, and such that all sensors are queried at least once 
during a cycle. 
After building this set, the base station must activate a subset of nodes at a 
time. In other words, only the sensor nodes from the active subset are activated 
during a time interval and all the others have their radios and sensors turned 
off in order to save energy and extend the WSN lifetime.
Simulations using real data show that such approach can successfully achieve 
its goals depending on the user requirements and on the quality of the data.
Similarly, our mechanism also assumes the task of selecting which sensors are 
going to be active in the next time interval. 
However, our mechanism is able to react to environmental changes, while their 
work is less dynamic. That is, once the sets of sensors are defined, they will 
be interleaved independently of changes that may happen around the WSN. 
We highlight that it may be possible to improve our mechanism by adopting their 
techniques to build the groups of sensors in a way that there is no reduction in 
the quality of the measurements and the energy savings are maximized.

The solution presented in \cite{Deshpande2004} (called BBQ) is a centralized 
mechanism used to query data based on sensor models. 
It assumes that the costs of retrieving data from many nodes can be extremely 
high and that sensors in close proximity are likely to have correlated readings, 
which may mean that most of the data provides little benefit in the quality of 
the answers given to the user.
In order to save energy, the BBQ incorporates statistical models of real-world 
processes into the query processing architecture and acquires data from the 
sensors only when the model itself is not sufficiently rich to answer the query 
with acceptable confidence.
To achieve such a goal, the BBQ approximates the probability density function 
of the measurements to multivariate Gaussian distributions and, given the 
correlation between the known measurement(s) and the unknown one(s), it 
calculates their expected value associated to a confidence interval. If the 
confidence level is greater or equal than a chosen threshold, it assumes that 
such value satisfies the system requirements. 
Otherwise, it calculates the energy consumed to retrieve new measurements 
considering the costs to activate the corresponding sensor and, finally, builds 
a query that will require the lowest energy consumption for the WSN and will 
give at least the minimum level of confidence set by the user.
Similarly to our mechanism, it exploits the correlation between different types 
of data that the sensor nodes may be able to measure, for example, their own 
voltage and the local temperature. The difference from our work is that they do 
not provide a method to measure the quality of the measurements and the 
performance of the system.

\section{Proposed Mechanism}
\label{sec:simulations}

Our system architecture is ready to use information from external WSNs, as 
described in~\cite{Pal2012} and~\cite{OechsnerVisions2014}.
To achieve the goal of optimizing the performance of the WSNs, they must be 
interconnected through their respective Enhanced Gateways (EGs). 
We explain the details of the mechanism in the following.



\subsection{Centralized Decisions}
\label{section:predictions}

\begin{figure}[h]
        \centering
        \begin{subfigure}[t]{0.21\textwidth}
                \centering
                \includegraphics[width=\textwidth]{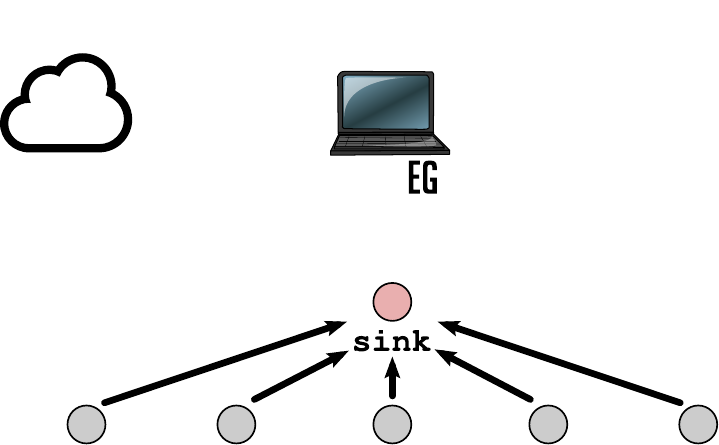}
                \caption{State 1: Nodes report measured values to the sink.}
                \label{fig:step-1}
        \end{subfigure}%
        \qquad
        \begin{subfigure}[t]{0.21\textwidth}
                \centering
                \includegraphics[width=\textwidth]{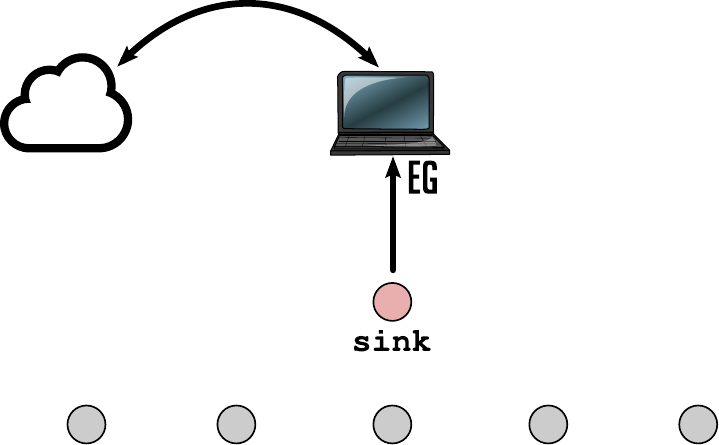}
                \caption{State 2: The sink transmits to the EG the sampled 
data, which 
                         can be forwarded to other EGs at the same time as they 
                         transmit to this WSN.}
                \label{fig:step-2}
        \end{subfigure}%
        \qquad
        \begin{subfigure}[t]{0.21\textwidth}
                \centering
                \includegraphics[width=\textwidth]{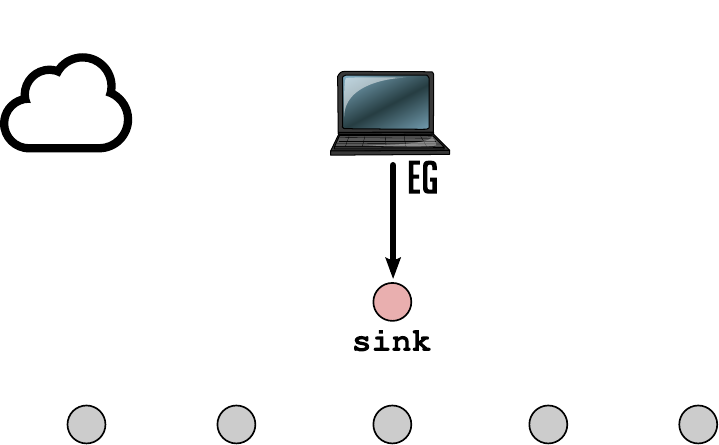}
                \caption{State 3: EG computes a new plan for the nodes and 
transmits the 
                         new configuration to its sink.}
                \label{fig:step-3}
        \end{subfigure}%
	\qquad
        \begin{subfigure}[t]{0.21\textwidth}
                \centering
                \includegraphics[width=\textwidth]{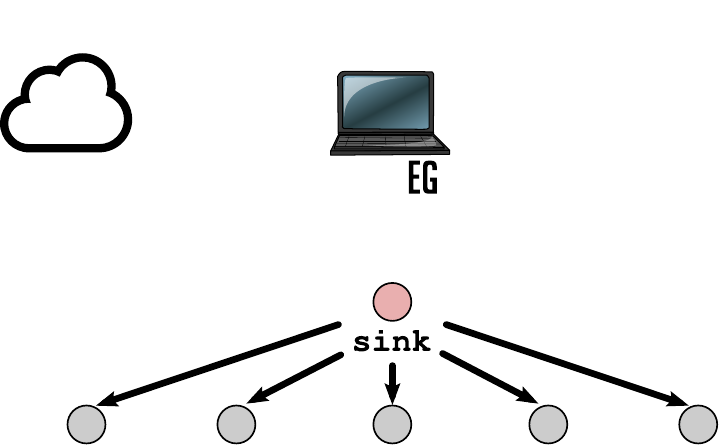}
                \caption{State 4: The sink updates the nodes' configuration.}
                \label{fig:step-4}
        \end{subfigure}%
        \caption{Different states of a WSNs using inter-WSN information 
exchange}
        \label{fig:steps}
\end{figure}

Periodically, the data retrieved by the nodes are transmitted to the sink. 
After receiving all the measurements, the sink computes
the received values before reporting them to the \EG{}, which may forward them 
to external WSNs. In parallel, the \EG{} may also receive information from 
external WSNs and, up to this point, all the data are collected and stored for 
further analysis. In intervals, the \EG{} uses the collected data to predict if 
there will be changes in the near future. Figure~\ref{fig:steps} describes the 
possible states of a WSN. 

The predictions done by the \EG{} can have two different outcomes: 
\emph{positive}, when changes in the environment are expected; and 
\emph{negative}, otherwise.
If an \EG{} receives information from internal and external sources, each 
prediction may be based on a different data type and independent for each 
metric. In such cases, they can be combined in order to produce only one 
outcome.
The outcomes can be compared with the real observations in order to verify the 
performance of the predictions. 
The feedback can be incorporated by the \EG{s} in order to improve their 
future decisions.

\subsection{Applications}
\label{section:approaches}

Based on the outcome of its prediction, a \EG{} selects the new strategy that 
the WSN must follow and will be applied by the sink. 
At the end, the sink transmits to its nodes a new configuration that they must 
follow in the next time interval, which may be an instruction to (de)activate 
themselves or to change the sensing intervals:

\subsubsection{Adaptive sensor nodes selection}

This application reduces the energy consumption of the network by deactivating 
some nodes during a certain period of time. In other words, when a node is 
deactivated, it does not make any measurement, but it may forward messages 
exchanged by their neighbors.
We recall that the sets of active nodes can follow the guidelines described 
in~\cite{Yann-Ael2005}, so the energy savings can be maximized without 
compromising the quality of the measurements.

\subsubsection{Adaptive sampling}

Differently from the other application, this solution does not change the 
number of active nodes. However, when the \EG{} has a \emph{positive} outcome 
and changes are expected in the environment, the nodes should reduce the 
time between two consecutive measurement transmissions, consuming more energy 
and producing more information about the environment. Otherwise, the energy can 
be saved, because it is not expected big changes in the environment.

\subsection{Quality of Measurements (QoM)}
\label{sec:qom}

As explained before, one of the goals of this mechanism is to reduce the 
energy consumed in a WSN without reducing the QoM (i.e., a parameter that 
evaluates if the gathered information from the environment during a certain 
period is enough to accurately represent it). However, the level of the QoM 
depends on the type of information reported by the nodes.

We consider monitoring WSNs that make continuous transmissions to the 
sink and tolerate a small number of packet losses as well as delays between 
consecutive transmissions, but do not allow the reduction of the covered area 
because it might miss changes occurring in certain subareas. Therefore, we 
scaled the QoM as shown in Table~\ref{table:qom}. There, each interval with a 
\emph{positive} outcome should be covered by more reports, increasing the level 
of knowledge about the environment. Although a high number of measurements 
always represents a \emph{good} QoM, the intervals with a \emph{negative} 
observation can be covered by less reports without compromising the quality, 
thereby saving energy. Periods with a \emph{negative} observation that are 
wrongly predicted mean that the system expected to have a \emph{positive} 
observation in them, produced more measurements and, thus, wasted energy. 
Differently from the states that a \emph{positive} is observed and the WSN 
produced a low number of measurements, those periods still have a \emph{good} 
QoM, but the energy consumption might have been reduced and the WSN lifetime 
increased.

\begin{table}[h]
\centering
  \hspace*{0.5cm}
 \def\arraystretch{2}
\begin{tabular}{cc|c|c|c}
\cline{3-4}
& & \multicolumn{2}{ c| }{\textbf{Prediction outcome}} \\ \cline{3-4}
& & \emph{positive} & \emph{negative} \\ \cline{1-4}
\multicolumn{1}{ |c 
}{\multirow{2}{*}{\begin{minipage}{0.8in}\centering\textbf{Actual \\ 
observation}\end{minipage}} } &
\multicolumn{1}{ |c| }{\emph{positive}} & \cellcolor[gray]{0.8} GOOD & BAD &    
 \\ 
\cline{2-4}
\multicolumn{1}{ |c  }{}                        &
\multicolumn{1}{ |c| }{\emph{negative}} & GOOD & \cellcolor[gray]{0.8} GOOD &   
  \\ 
\cline{1-4}
\end{tabular}
\caption{Definition of QoM.}
\label{table:qom}
\end{table}

Based on this, the accuracy was defined as the percentage of intervals in a day 
in which the system was operating in a highlighted state. Moreover, the accuracy 
of \emph{positives} is the percentage of intervals with \emph{positives} 
covered by a high number of measurements.

Regarding the system operation, during intervals in which 
variations are predicted and the predictions have \emph{positive} outcomes, 
the \EG{} updates the operation of its WSN in order to collect more 
information. 
Each update on its operation affects either the number of active nodes or the 
time interval between two measurements done by the sensors. As a consequence of 
this, the number of measurements, the number of transmissions and the energy 
consumption have higher values during these periods of time, while the opposite 
effect occurs when no variation is predicted.

\subsection{Performance score}
\label{sec:performance}

In order to evaluate how efficient the use of external information can be, we 
developed a way to compare the approaches. For a given scenario, we calculate 
the lowest energy consumption that the WSN may have ($E_\text{min}$), which 
can be done by always setting the plan that produces less measurements during a 
day. On the other hand, we measure how much energy is consumed by the WSN when 
it produces the maximum number of measurements during the same time interval 
($E_\text{max}$). Thus, the percentage of energy saved by an approach 
($E_\text{ps}$) is derived from the energy consumed ($E_\text{consumed}$) by 
the 
relation:

\begin{equation}
E_\text{ps} = \frac{E_\text{max} - E_\text{consumed}}{E_\text{max} - 
E_\text{min}}
\end{equation}

A correct prediction about a \emph{negative} observation means that the 
system is producing less measurements and saving energy. Therefore, this 
accuracy factor is implicitly inserted in the value of $E_\text{ps}$ and should 
not be considered again in the final equation. Considering this, the trade-off 
between the QoM and the energy consumption can be calculated if we use only the 
percentage of predictions of \emph{positive} outcomes ($P_\text{\highDelta{}}$) 
that the system could successfully do:

\begin{equation}
P_\text{\highDelta{}} = \frac{\text{\# of positives correctly 
predicted}}{\text{\# of observed positives}}
\end{equation}

Finally, the \emph{Performance score} ($p$) is defined as the product 
between the percentage of saved energy and the percentage of \emph{positives} 
correctly predicted, which quantifies how much the system actually 
consumes to have such level of accuracy. If interpreted as a dot product 
between two vectors, the highest value represents the system having the highest 
possible energy savings and the highest possible accuracy \highDelta{}s:

\begin{equation}
p_{(\alpha)} =  {{E_\text{ps}}^\alpha } \cdot { 
{P_\text{\highDelta{}}}^{(1 - \alpha)} }
\label{eq:performance}
\end{equation}
where $0 \leq \alpha \leq 1$ is the exponent that represents the system's 
priority on the energy saved over its accuracy. For example, if $\alpha < 0.5$, 
the energy savings will have a bigger impact at the performance score. 
Obviously, if $\alpha = 0.5$, the system will not prioritize any of them. 

\section{Use Case}
\label{sec:current-simulations-scenario}

To create a realistic use case, we used the temperature and relative humidity 
of $16$ days measured by three different nodes in the experiments done 
in~\cite{Yann-Ael2005}. 
The simulated use case is based on a real scenario from where the data was 
fetched: an office with two WSNs deployed close to each other. 
There, nodes are positioned in a grid topology with two different WSNs: 
\emph{Network A} monitoring temperature and \emph{Network B} monitoring 
relative humidity.

\emph{Network A} has one node that retrieves data from the environment, and a 
sink node that receives the temperature values and transmits them to the 
respective EG (\EG{A}), which forwards everything to \EG{B}. On the other side, 
\emph{Network B} was composed by $26$ nodes that monitor the relative humidity 
plus a sink connected to \EG{B}, which is responsible for averaging the values 
received after each measurement. Based on the data received from \EG{A} and on 
the stored averages, \EG{B} is able to set different WSN operation plans, and to 
communicate the required changes to its sink node in order to forward them to 
the wireless sensor nodes.

\subsubsection{Adaptive sensor nodes selection}

We manually created three different sets of active nodes for the \emph{Network 
B}: One with half of the nodes plus the sink; another with the other half plus 
the sink; and the last one with all nodes together. The first two plans are 
used for saving energy and are switched on every update to extend the WSN's 
lifetime, while the goal of the all-nodes plan is to provide more information 
about the environment. The downside is that this plan consumes more energy. 
Therefore, the latter is only used when the prediction produce 
\emph{positive} outcomes and the environment is expected to change. 

\subsubsection{Adaptive sampling}

When the prediction outcome is a \emph{positive} and changes are expected in 
\emph{Network B}, nodes take measurements and transmit them every $30$ seconds, 
consuming more energy and producing more information about the environment. 
Otherwise, this is done every $180$ seconds.

\subsection{Constant Predictions}
\label{sec:system-goal-definition}

At runtime, \emph{Network B} defines how its nodes will react to environmental 
changes based on the predictions done: reporting more information when the 
environment is supposed to undergo variations and saving energy otherwise.
In order to predict these variations, we calculated the average of the 
temperature and relative humidity values, without mixing data types, in 
discrete and sequential $5$-minute window intervals. The absolute difference 
between the averages of two consecutive intervals is denoted \ourDelta{}.
In order to identify the data types, we used subscripts: \ourDelta{T} 
for temperature values and \ourDelta{RH} for relative humidity values.
We have assumed that a large difference between the averages 
represent significant changes in the environment.
Therefore, the system goal is to predict whether the next \ourDelta{} will be 
over a determined threshold, \threshold{}, or not. 
To achieve that, we used a constant na\"ive model to make the predictions, 
i.e., in case of $\ourDelta{} > \threshold{}$, we label it as \highDelta{}, 
representing a \emph{positive} outcome; 
otherwise, we call it a \lowDelta{}. 

\begin{figure}[h]
	\centering
	\begin{subfigure}[t]{0.21\textwidth}
		\centering
		
\includegraphics[width=\textwidth]{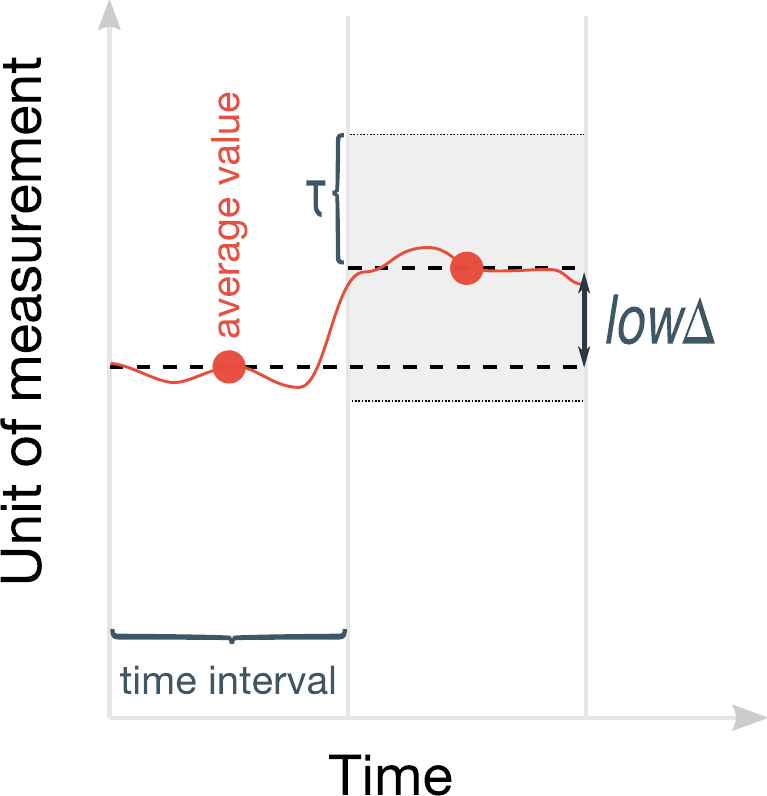}
		\caption{The concept of \lowDelta{}}
		\label{fig:delta-low-plus}
	\end{subfigure}%
	\qquad
	\begin{subfigure}[t]{0.21\textwidth}
		\centering
		
\includegraphics[width=\textwidth]{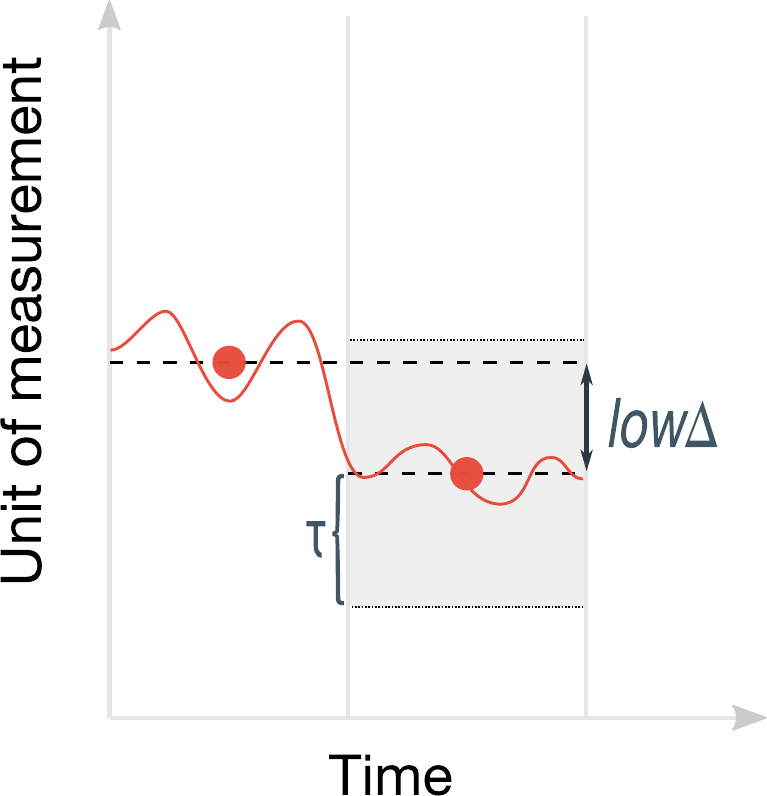}
		\caption{Another case of \lowDelta{}}
		\label{fig:delta-low-minus}
	\end{subfigure}%
	\qquad
	\begin{subfigure}[t]{0.21\textwidth}
		\centering
		
\includegraphics[width=\textwidth]{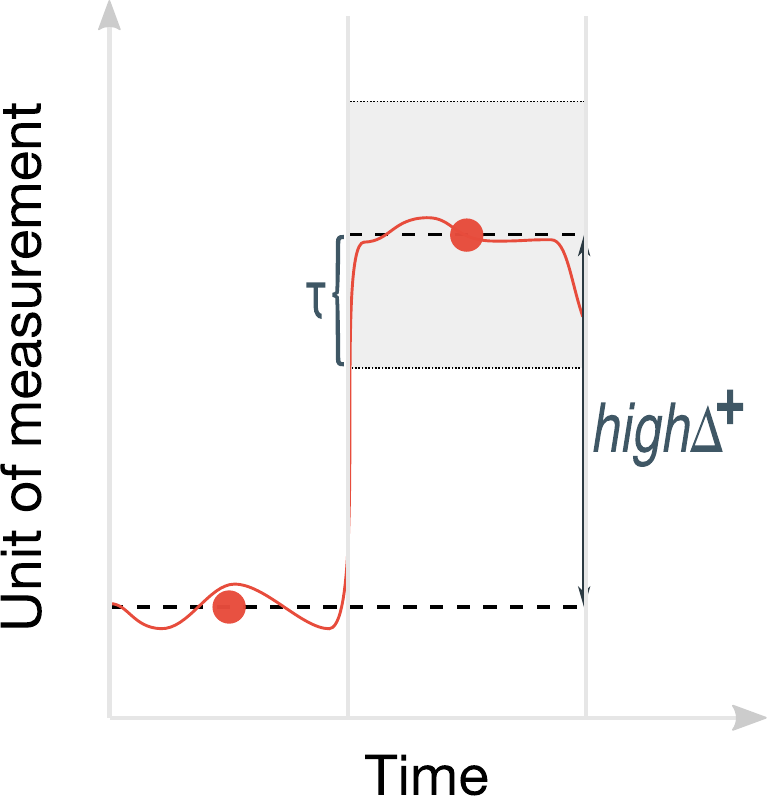}
		\caption{The concept of \highDelta[+]{}}
		\label{fig:delta-high-plus}
	\end{subfigure}%
	\qquad
	\begin{subfigure}[t]{0.21\textwidth}
		\centering
		
\includegraphics[width=\textwidth]{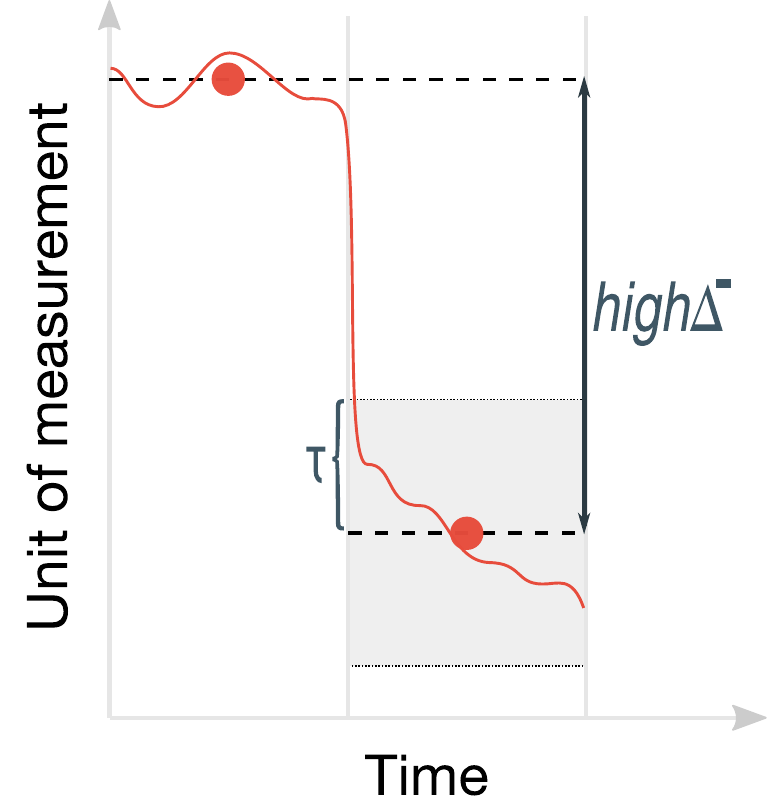}
		\caption{The concept of \highDelta[-]{}}
		\label{fig:delta-high-minus}
	\end{subfigure}%
	\caption{How the system labels the \ourDelta{}s.}
	\label{fig:delta}

\end{figure}

In some cases, it may be useful to know if a \highDelta{} means that the 
average is increasing or decreasing. In order to identify it, we added an 
additional notation to \ourDelta{}. If the most recent average computed 
differs more than \threshold{} and is greater 
than the penultimate one, we mark it as \highDelta[+]{}; 
if it differs more than \threshold{} but is lower, we use 
\highDelta[-]{}, as shown in Figure~\ref{fig:delta}.
In case of having a \lowDelta{}, there is no need for highlighting if the value 
is greater or less than the penultimate one.

Predictions are independent for each metric.
Furthermore, any prediction is 
composed by three factors: the last two symptoms and the last prediction. The 
general idea is to try to learn the trend and avoid wrong predictions provoked 
by noise and outliers. Thus, every time that two factors agree in one 
direction, the prediction is that, in the next interval, the environment will 
follow it. 
Otherwise, if the three factors are different, the prediction is that the 
environment will not undergo variations in the near future. 
Table \ref{table:system-states} shows how we did the predictions using
\ourDelta{}s.

\begin{table}[h]
 \centering
\begin{tabular}{cccc}
\toprule
\multicolumn{2}{c}{\textbf{Last Symptoms}} & \textbf{Last Prediction} & 
\textbf{Prediction} \\ \midrule

\lowDelta{}         & \lowDelta{}            & any                  
& 
\lowDelta{}           \\ \midrule

\highDelta[+]{}     & \highDelta[+]{}    
  & any                  & \highDelta[+]{} \\ 
\midrule
\highDelta[-]{}     & \highDelta[-]{}    
  & any                  & \highDelta[-]{} \\ 
\midrule

\highDelta[+]{}     & any            & 
\highDelta[+]{} & 
\highDelta[+]{} \\ \midrule

\highDelta[-]{}     & any            & 
\highDelta[-]{} & 
\highDelta[-]{}                 \\ \midrule

\lowDelta{} & any           & \lowDelta{}        
         & \lowDelta{} \\ \midrule

\highDelta[+]{}     & \highDelta[-]{}        
& \lowDelta{}                 & \lowDelta{}                 \\ 
\midrule
         \highDelta[+]{}     & \lowDelta{}            & 
\highDelta[-]{} & \lowDelta{}                 \\ \midrule
\highDelta[-]{}     & \lowDelta{}            & 
\highDelta[+]{} & \lowDelta{}                 \\ \bottomrule
\end{tabular}
\caption{How the system reacts to the symptoms.}
\label{table:system-states}
\end{table}

Finally, if a \emph{EG} receives information from internal and external 
sources, each prediction may be based on a different data type.  In this case, 
it combines them in the simplest way: if one of the predictions is labeled as 
\highDelta{}, the final prediction is a \highDelta{}; otherwise, it is a 
\lowDelta{}.

\subsubsection{Adaptive threshold}

The value of \threshold{} is set based on the proportion of \ourDelta{}s seen 
in the historical data. For example, if the goal is to predict the highest 
quarter of \ourDelta{}s in a day, the threshold will be set at the $75$th 
percentile of \ourDelta{}s. In this case, we identify it with the number $75$ 
subscripted: \threshold[75].

\subsubsection{Symptoms}
\label{sec:symptom-definition}

To make those predictions, we must observe the measurements and find 
symptoms. A symptom, \symptom{}, is defined as a value where a 
$\ourDelta{} > \symptom{}$ represents a high probability of having 
$\ourDelta{} > \threshold{}$ in the next interval. Therefore, if we 
notice that the most recent \ourDelta{} is greater than \symptom{}, we have a 
symptom of \highDelta{}; otherwise, it is a symptom of \lowDelta{}.
Even though the concepts of \symptom{} and \threshold{} are similar, the 
numerical values may be different. For example, after observing the historical 
data, we might notice that every $\ourDelta{} > \threshold[40]$ calculated at 
time $t$ was followed by a $\ourDelta{} > \threshold[75]$ at time $t + 1$. So, 
we would set the value of \symptom{} at the $40$th percentile of \ourDelta{}s.


\section{Evaluation}
\label{sec:evaluation}

We considered each measurement done by the real nodes as the average of the 
network measurements in our simulations. 
Moreover, each set of measurements done by a node in a day was considered one 
day's worth of data. Therefore, we had enough data to simulate $48$ different 
days.
To check the feasibility of using this solution in the presented 
scenario, we evaluated the energy consumption in OMNeT++~\cite{Varga2001} and 
the calculations about the performance score in Matlab.
First, using OMNeT++ and MiXiM~\cite{Kopke}, we simulated the energy 
consumption based on TelosB nodes~\cite{Platform} using BMAC~\cite{Fakih2006} as 
MAC protocol and a flooding routing protocol. 
In these simulations, the sensor nodes received new plans from the \EG{} every 
$5$ minutes, as explained in Section~\ref{section:approaches}. We calculated 
the average energy consumption on each plan, considering also the energy spent 
to disseminate the plan changes through the network. 

In Matlab, the data from the sensors were split into a training and a validation 
datasets to avoid overfitting. Each of these datasets was defined by a set of 
$24$ days that were randomly selected on each run (repeated random sub-sampling 
validation). The model was fit to the training data, and predictive accuracy was 
assessed using the validation data. The tests were done over $10$ different 
combinations of days and the final results were averaged over the splits. In the 
end, we checked how the system behaved when the plan of \emph{Network B} was 
selected using only internal information (relative humidity values), only 
external information (temperature values) and combining both, and used the 
energy consumption levels to plot the results. 

\subsection{Training dataset}

\begin{figure}[h]
	\centering
	\begin{subfigure}[t]{0.21\textwidth}
		\centering
		
\includegraphics[width=\textwidth]{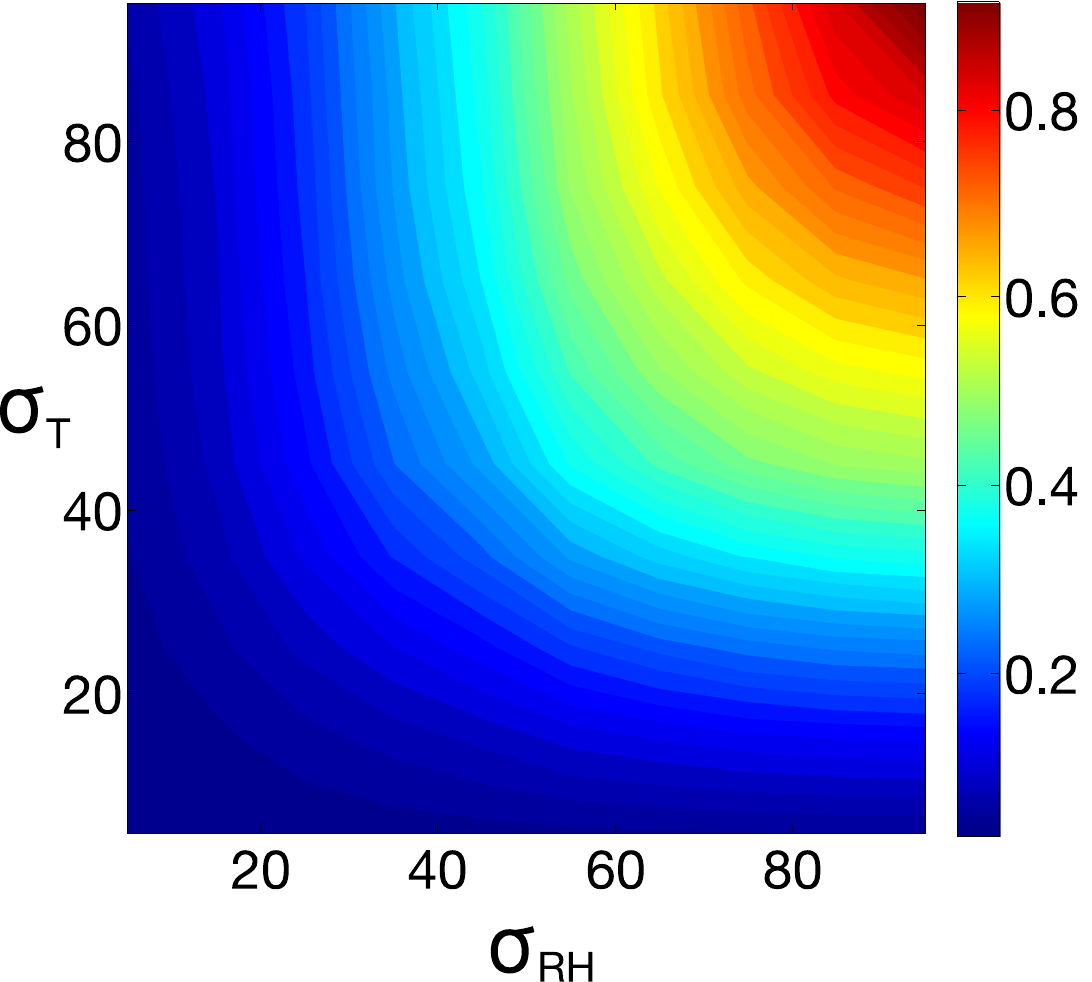}
		\caption{Energy saved ($E_\text{ps}$)}
		\label{fig:performance-energy-consumption}
	\end{subfigure}%
	\qquad
	\begin{subfigure}[t]{0.21\textwidth}
		\centering
\includegraphics[width=\textwidth]{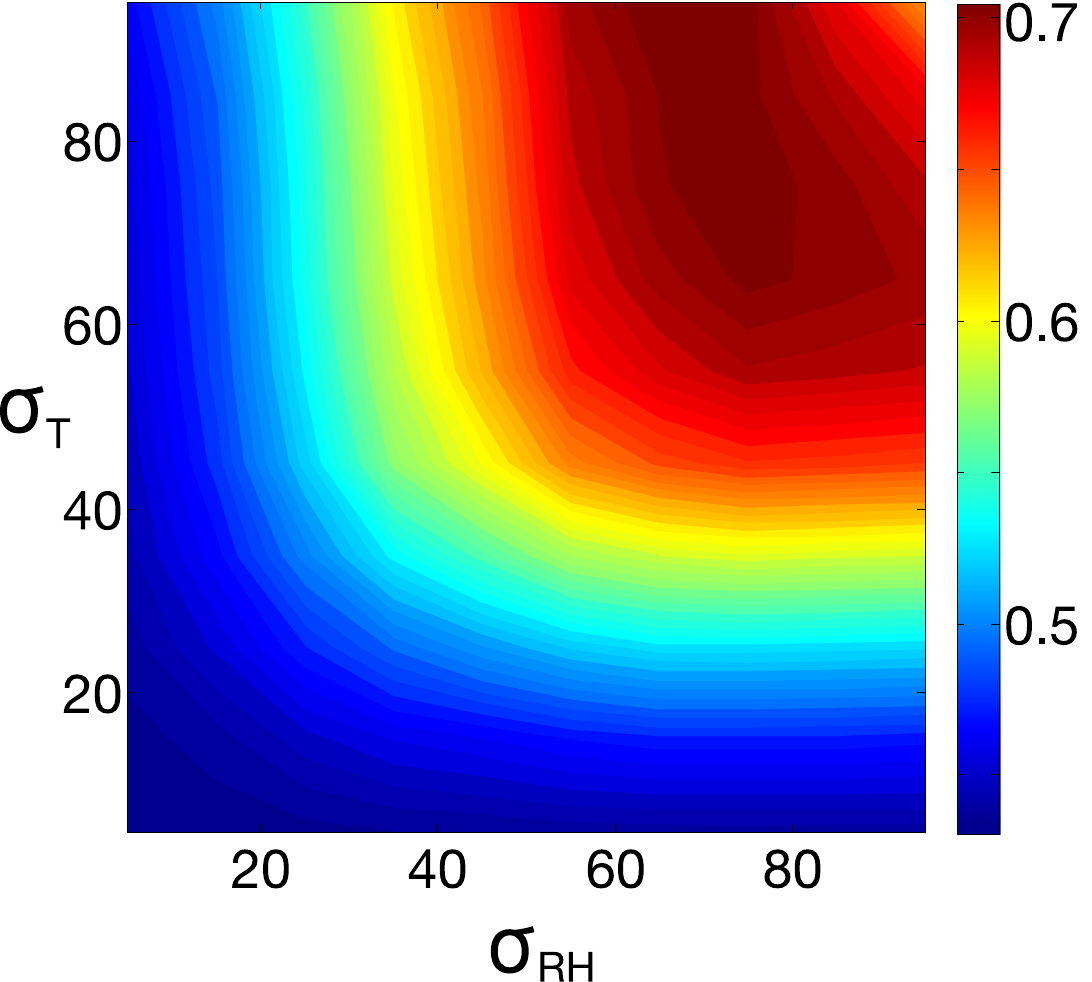}
		\caption{Accuracy of the predictions}
		\label{fig:performance-accuracy}
	\end{subfigure}%
	\qquad
	\begin{subfigure}[t]{0.21\textwidth}
		\centering
		
\includegraphics[width=\textwidth]{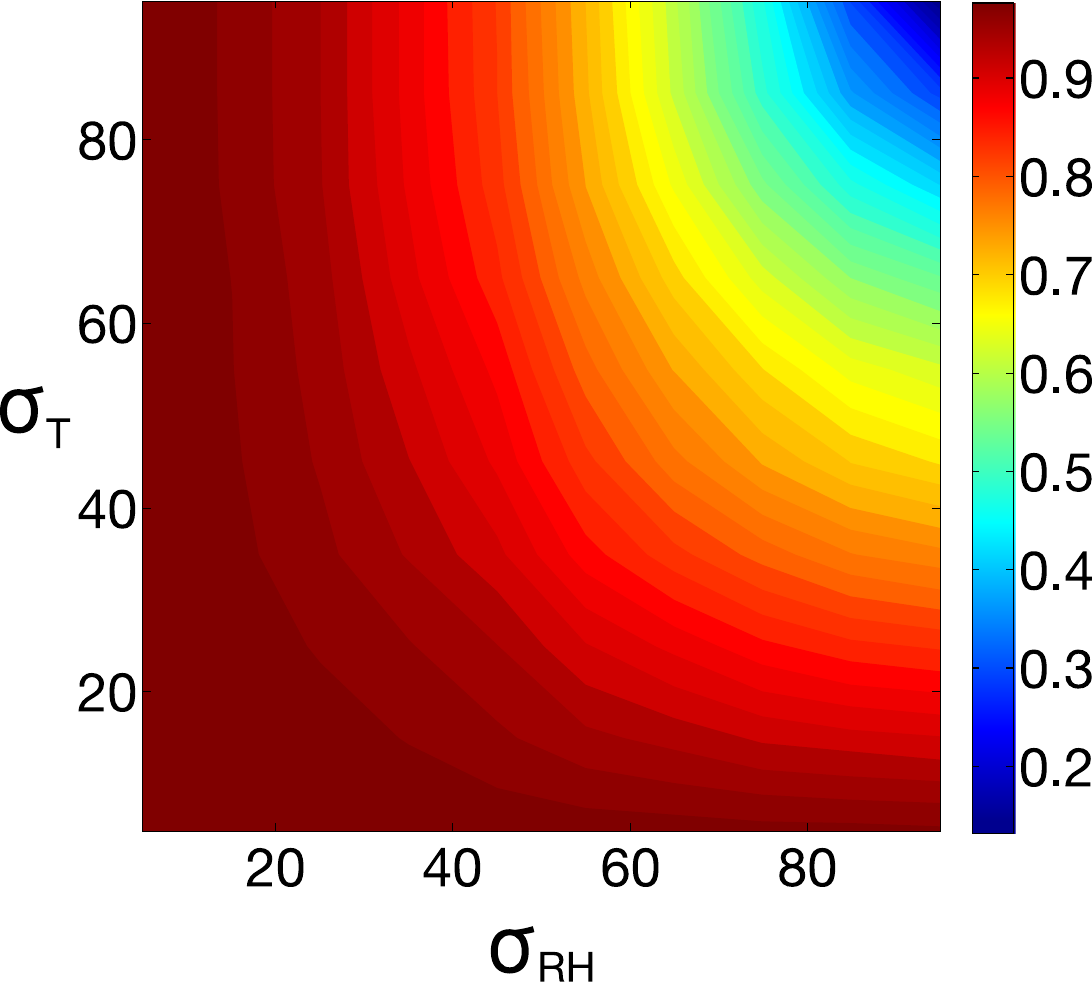}
		\caption{Accuracy of \highDelta{}s ($P_\text{\highDelta{}}$)}
		\label{fig:performance-high-accuracy}
	\end{subfigure}%
	\qquad
	\begin{subfigure}[t]{0.21\textwidth}
		\centering
		
\includegraphics[width=\textwidth]{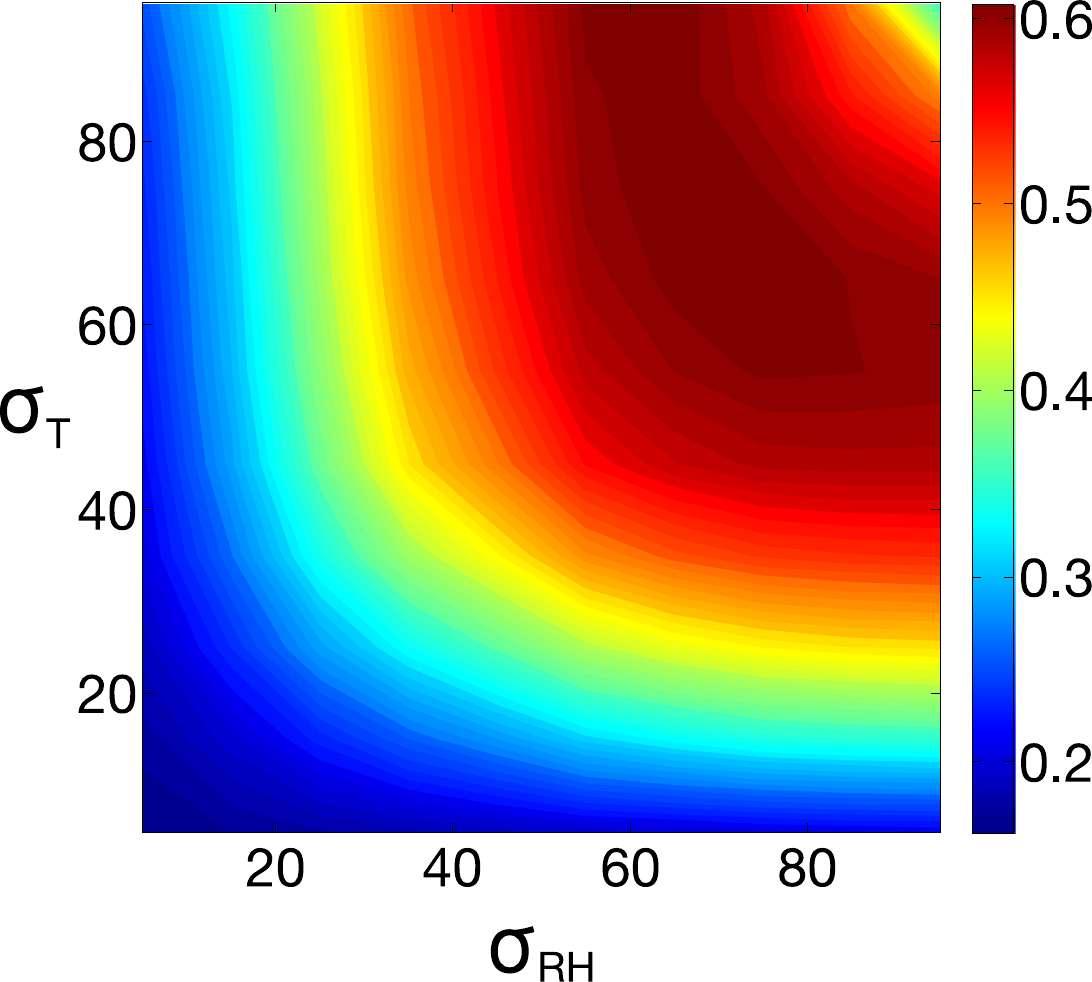}
		\caption{$p_{(0.5)}$}
		\label{fig:performance-score}
	\end{subfigure}%
	\caption{Parameters obtained using the training data.}
	\label{fig:performance-process}

\end{figure}

After selecting $24$ days for the training dataset, the measured values were 
used to set three different parameters: 

%
\begin{itemize}
 \item \textbf{The value of \threshold{}} -- The threshold that the 
\EG{}s must set. It was calculated as explained in 
\ref{sec:system-goal-definition}, based on the measurements done during the 
training days.
 \item \textbf{The values of \symptom{}s} -- The system built a table with the 
values of $p$ based on percentiles, as shown in 
Figure~\ref{fig:performance-score}. The 
numerical value of \symptom[T] and \symptom[RH] was the same as the 
percentiles of \ourDelta{T} and \ourDelta{RH}  with the highest value 
of $p$.
\end{itemize}

We assumed that the saved energy and the accuracy of the system have similar 
importance and set the value of $\alpha = 0.5$ in Equation~\ref{eq:performance}.
Figure~\ref{fig:performance-energy-consumption} shows how much energy can be 
saved based on the thresholds that are used as symptoms of future changes. For 
example, at the point $(40, 20)$, any \ourDelta{RH} over the $40^\text{th}$ 
percentile (i.e., greater than $40\%$ of the values) is considered as a symptom 
of change, as well as any \ourDelta{T} over the $20^\text{th}$ percentile. When 
a symptom is detected, the EG may launch a plan to produce more measurements in 
the next time-interval and, consequently, consume more energy.
Figure~\ref{fig:performance-accuracy} shows the total accuracy of the 
predictions and Figure~\ref{fig:performance-high-accuracy} shows how the 
accuracy of \highDelta{}s changes depending on the threshold chosen to represent 
a symptom of changes in the future.

\subsection{Validating dataset}

The other $24$ days were considered part of the validating dataset and their 
data were used to validate whether the system had chosen well and whether our 
hypothesis was valid. For this, the system used all the parameters 
calculated in the last step to calculate~$p$.


\subsection{Results}

\begin{figure}[h]
	\centering
%
	\begin{subfigure}[t]{0.4\textwidth}
		\centering
		
\includegraphics[width=\textwidth]{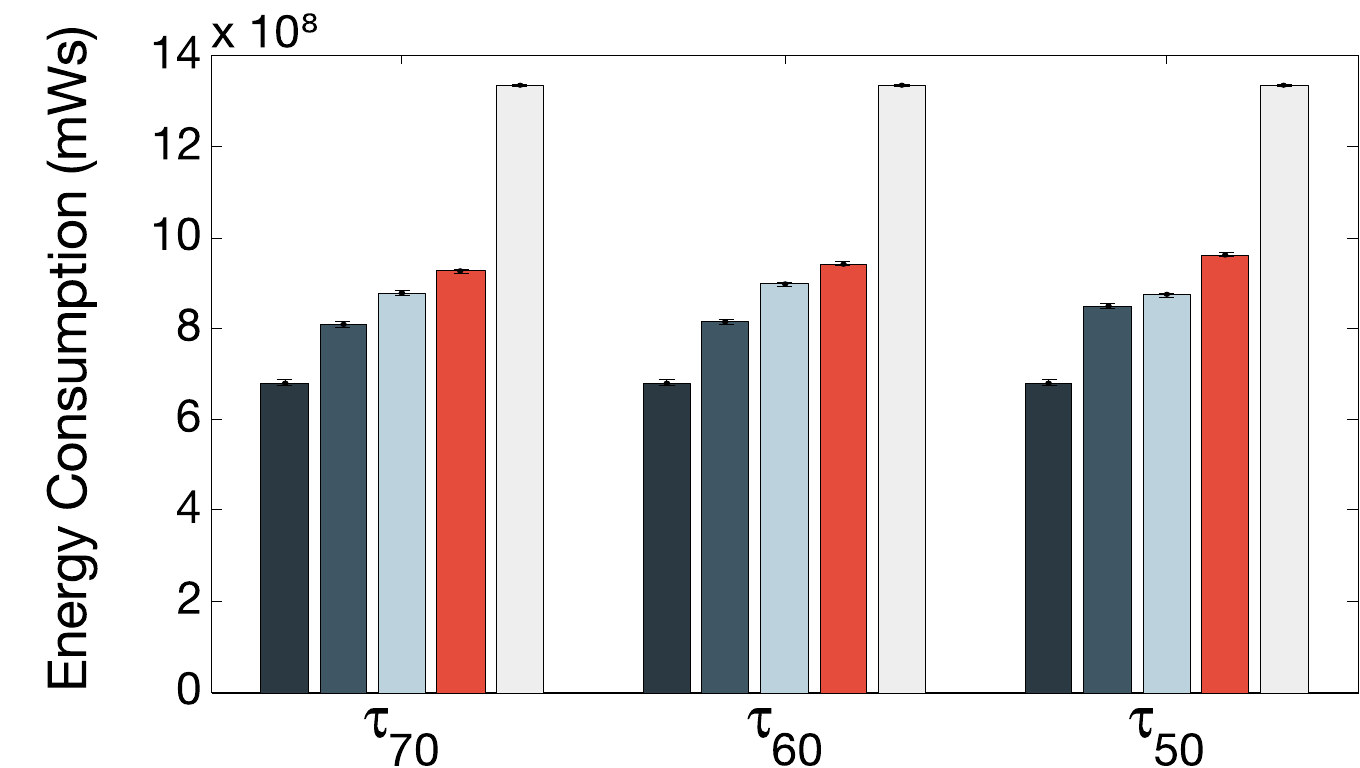}
		\caption{Adaptive sampling.}
		\label{fig:results-energy-2}
	\end{subfigure}%
	\qquad
	\begin{subfigure}[t]{0.4\textwidth}
		\centering
		
\includegraphics[width=\textwidth]{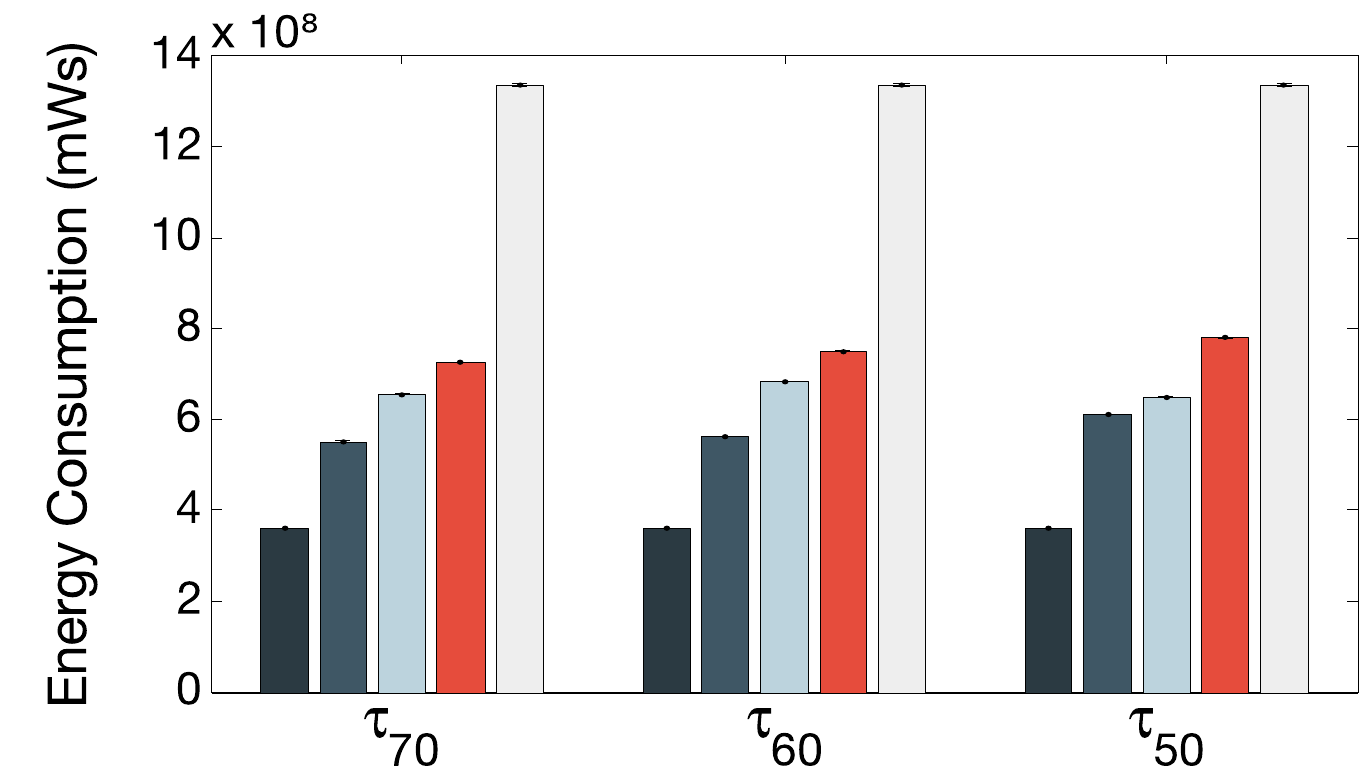}
		\caption{Adaptive sensor nodes selection.}
		\label{fig:results-energy-3}
	\end{subfigure}%
	\qquad
	\begin{subfigure}[t]{0.4\textwidth}
		\centering
		
\includegraphics[width=\textwidth]{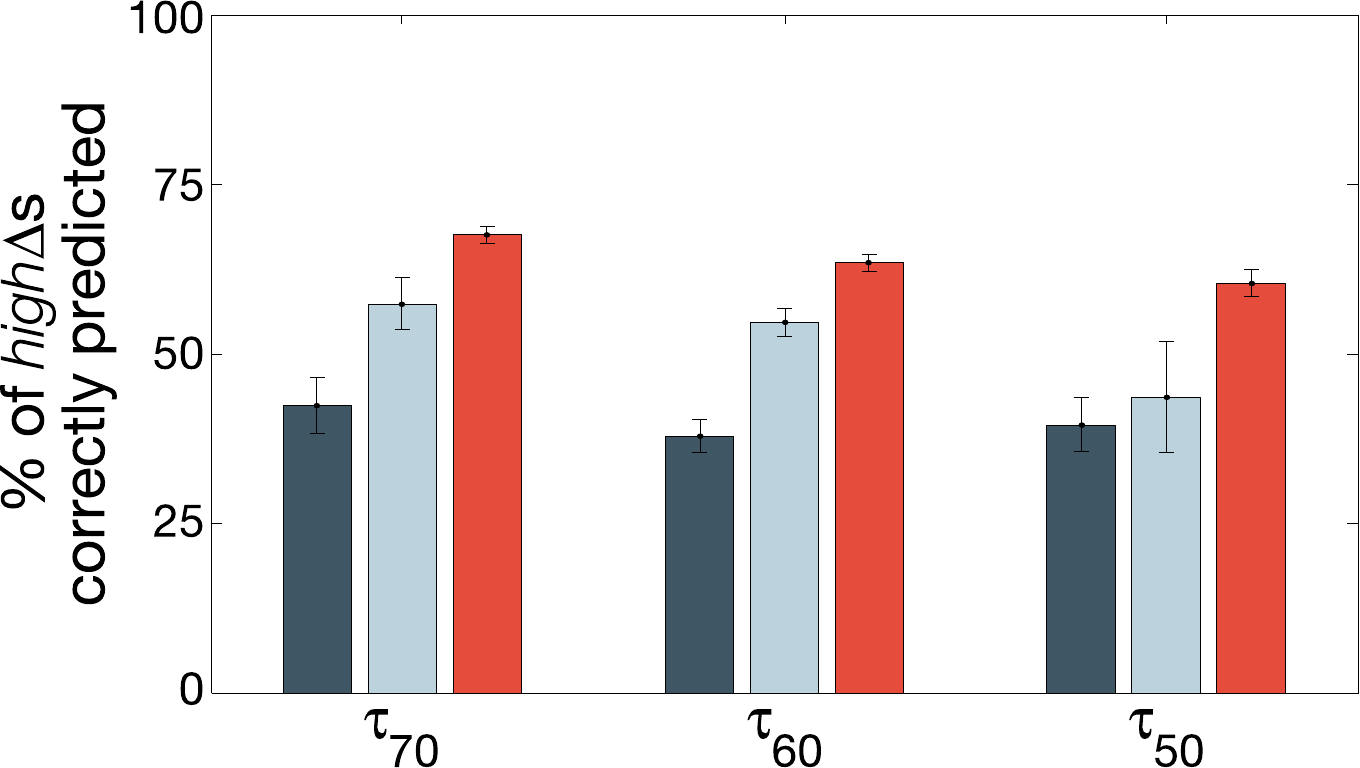}
		\caption{Accuracy of the predictions of \highDelta{}s.}
		\label{fig:results-accuracy_high}
	\end{subfigure}%
	\qquad
	\begin{subfigure}[t]{0.4\textwidth}
		\centering
		
\includegraphics[width=\textwidth]{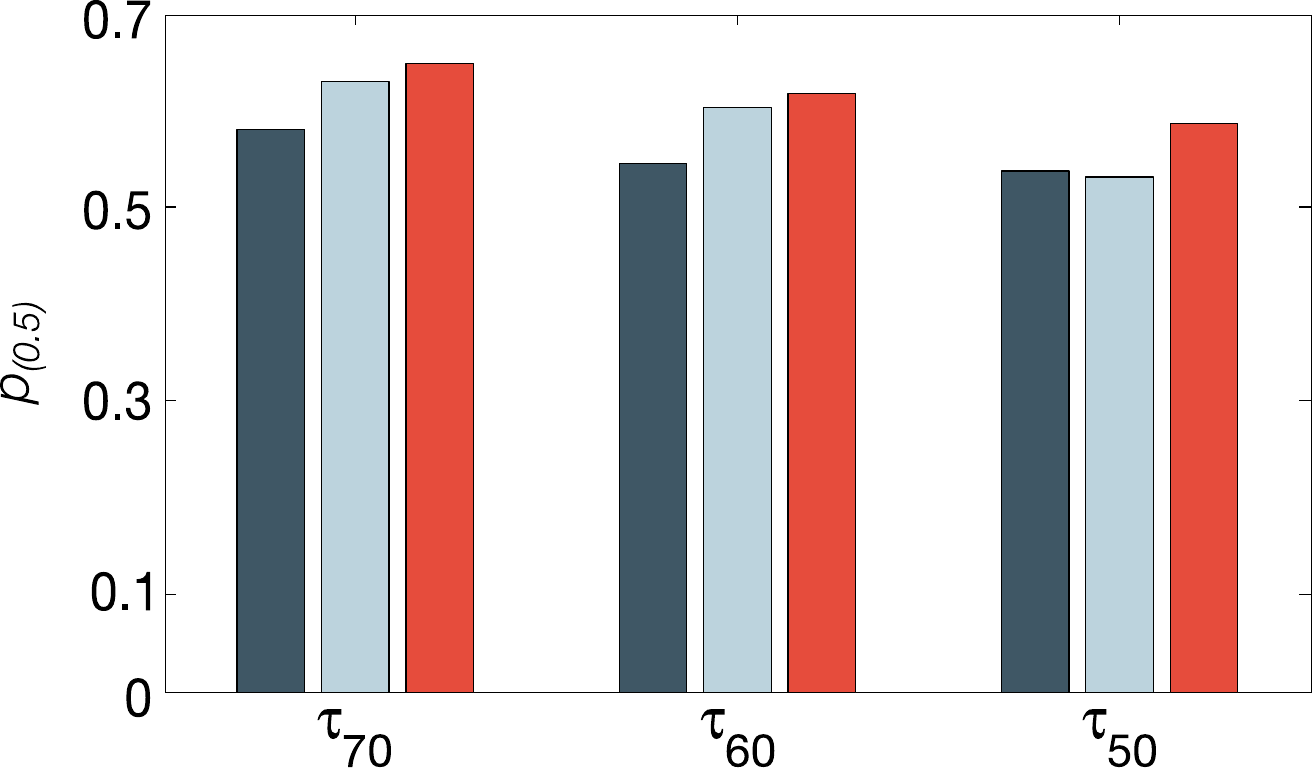}
		\caption{Performance score of each approach.}
		\label{fig:results-eval}
	\end{subfigure}%
	
	\centering
	\includegraphics[width=0.9\textwidth]{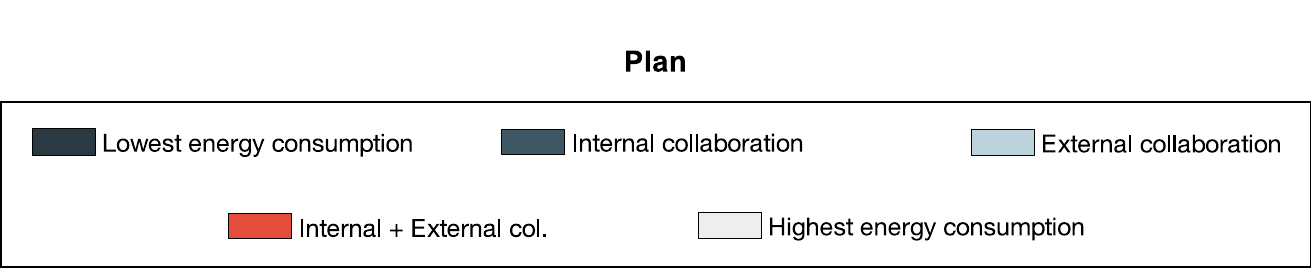}
	\label{fig:results-legend}
	
	\caption{Simulation results.}
	\label{fig:results}
	
\end{figure}

The plots in Figure~\ref{fig:results} show the obtained results, where it is 
possible to see how much our solution was able to exploit the trade-off between 
the energy consumption and the quality of the measurements. To show better its 
benefits, we included two baseline scenarios that did not use collaboration: the 
first one saved the maximum energy possible by transmitting less measurements; 
the second did not save energy and always used the plan that transmits more 
measurements. An important remark is that both scenarios have $p_{(\alpha)} = 0$ 
for any $\alpha$, because either they did not save any energy (the highest 
consumption plan case) or their accuracy of detecting \highDelta{}s was zero 
(the lowest consumption plan case).

The results are split into three groups, according to the \threshold{} set for 
each case (\threshold[70], \threshold[60] and \threshold[50]). Each bar 
represents an average for the $24$ days of the validation dataset. Observing the 
data, we can see that the correlation between temperature and relative humidity 
values is closer to $-1$ when we consider only the highest \ourDelta{}s, i.e., 
\threshold[70]. Therefore, we assume that there are other factors that may 
influence the small variations in the relative humidity, such as the presence of 
persons close to the sensors. This explains why the percentage of \highDelta{}s 
correctly predicted is lower when the system tries to track a higher number of 
changes (\threshold[50]).


In Figures~\ref{fig:results-energy-2} and \ref{fig:results-energy-3}, we 
can observe that, when we used only the plan that changed the number of active 
nodes, the system spent around $54\%$ of the energy compared to the scenario in 
which the network was always producing more measurements. Also, 
Figure~\ref{fig:results-accuracy_high} shows that predictions can successfully 
improve the WSNs' operation. It is possible to see that, using only the relative 
humidity values as a reference (absence of external collaboration), $42.3\%$ of 
the $5$-minute intervals with \highDelta{RH}s were correctly predicted with 
\threshold[60]. Compared to that, we can observe that the energy consumption 
increased much less than the accuracy levels. For example, with \threshold[60], 
using the combination of internal and external information, the system was able 
to correctly predict $67.9\%$ more \highDelta{}s consuming only $33.5\%$ more 
energy. This means that the energy was used more intelligently in the second 
case.

Figure~\ref{fig:results-eval} shows that our approach for inter-WSN information 
exchange outperforms the other types of collaboration that use less information 
and spend their energy less efficiently. In summary, the trade-off between 
energy consumption and QoM was achieved and found to produce more effective 
results than the other approaches.




\section{Conclusion and Future Work}
\label{sec:conclusion}

Based on the presented results, it is possible to determine that our mechanism 
is able to use internal and external information to optimize the WSNs' 
performance, which is illustrated by the difference in the values of $p$. During 
the tests, we have also noticed that these improvements could be achieved only 
with data that is not only highly correlated, but there must also be a relation 
of causation between them. 
In this case, we noticed that changes in temperature led to changes in relative 
humidity, but the opposite was not necessarily true. Therefore, it would be 
more complex to make good predictions if we tried to predict 
temperature changes based on relative humidity values.

Although we made use of real data from existing experiments, we did generic 
calculations and assumptions that can be extended to numerous scenarios, in 
order to prove the general idea of this concept. We expect that specific 
knowledge about different scenarios may lead to better results. For example, as 
shown in \cite{Lawrence2005}, when the relative humidity is over $50\%$, it is 
possible to calculate its value based on information about the temperature 
only. Thus, in a scenario similar to ours, the system could save 
even more energy by letting the EG calculate the local data based on 
external information.


The next steps include adapt this solution to an autonomic system, as described 
in~\cite{6583430}. That is, a more generic mechanism which is able to work with 
other WSN types and is able to work with other prediction methods that may have 
better performance in different scenarios.
Additionally, the idea of an autonomic solution involves a pro-active and 
self-managing system, which improves the information fusion and the decision 
optimization, besides creating specific plans for the WSNs according to the 
predictions about the near future.




\section*{Acknowledgment}
This work has been partially supported by the Spanish Government through the 
project TEC2012-32354 (Plan Nacional I+D), by the Catalan Government 
through the project SGR2009\#00617 and by the European Union through the 
project FP7-SME-2013-605073-ENTOMATIC.



%

\bibliographystyle{unsrt}

\bibliography{bibliography}

\end{document}